\documentclass[12pts,epsf,aip,jap]{revtex4}
\usepackage{float}
\usepackage{color,graphicx}
\usepackage{hyperref}

\begin{document}

\title{Vortex dynamics and second magnetization peak in PrFeAsO$_{0.60}$F$_{0.12}$ superconductor}%
\author{D. Bhoi}
\author{P. Mandal}%
 \email{prabhat.mandal@saha.ac.in}
\affiliation{Saha Institute of Nuclear Physics, 1/AF Bidhannagar,Calcutta 700 064, India%\\This line break forced with \textbackslash\textbackslash
}%
\author{P. Choudhury}

\affiliation{Central Glass and Ceramic Research Institute, 196 Raja S. C. Mullick Road, Calcutta  700 032, India%\\This line break forced% with \\
}%

\date{\today}%

\begin{abstract}
We have studied the vortex dynamics in the PrFeAsO$_{0.60}$F$_{0.12}$ superconducting sample by dc magnetization and dynamic magnetization-relaxation rate $(Q)$ measurements. The field dependence of the superconducting irreversible magnetization $M_s$ reveals a second magnetization peak or fishtail effect. The large value of $Q$ is an indication of moderate vortex motion and relatively weak pinning energy. Data analysis based on the generalized inversion scheme suggests that the vortex dynamics can be described by the collective pinning model. The temperature dependence of the critical current is consistent with the pinning due to the spatial variation in the mean free path near a lattice defect ($\delta l$ pinning). The temperature and field dependence of $Q$ indicates a crossover from elastic to plastic vortex creep with increasing temperature and magnetic field. Finally, we have constructed the vortex phase diagram based on the present data.
\end{abstract}

\maketitle

\section{Introduction}

The relatively high superconducting transition temperature $T_c$ up to 55 K and phenomenally high upper critical field $H_{c2}$ along with lower anisotropy and larger coherence length compared to the high-$T_c$ cuprates have made the pnictides interesting candidates for possible technological applications \cite{john,stewart}. Another important feature of these compounds is the presence of a peak in the field dependence of critical current density, which is associated with a fishtail or second magnetization peak (SMP) effect in the $M(H)$ curve, as observed in several cuprates and conventional superconductors \cite{daeu,kel,gil,gil1,deli,khay,abu,sbhat,klein, klein1}. The appearance of SMP in the $M(H)$ curve is not a universal phenomenon in pnictides. Whether in a particular material the SMP would appear is very much sample specific \cite{shen,proz,habe,salem}. Most of the theoretical approaches agree that the temperature dependence of the field $H_p$, at which the SMP has its maximum, is related to the nature of vortex pinning and corresponds to a crossover between two different regimes of vortex lattice. Even though there are a lot of studies dealing with this phenomenon, the mechanism and origin of this effect are still much debated partly because the proposed models are system specific. \\

So far, the studies of vortex dynamics have been mainly focused onto the doped 122 family $A$Fe$_2$As$_2$ ($A$ = Ca, Ba, Eu, etc.) as sizeable single crystals are available in this group of pnictides. In the 122 compounds such as BaFe$_{2-x}$Co$_x$As$_2$ \cite{shen,proz}, Ca$_{1-x}$Na$_x$Fe$_2$As$_2$ \cite{habe}, and Ba$_{1-x}$K$_x$Fe$_2$As$_2$ \cite{salem} the vortex dynamics has been described by the collective pinning model. In these compounds, the SMP in the $M(H)$ curve signifies a crossover from elastic to plastic vortex creep regime. In contrast, studies of Bitter decoration, small-angle neutron scattering and magnetic force microscopy in 122 compounds have revealed highly disordered vortex-glass phase with a short-range hexagonal order \cite{eskil,inos,vinni}. Furthermore, these studies suggest that the vortices remain in the single-vortex pinning limit even at high-magnetic fields up to 9 T. Kopeliansky \emph{et al.} \cite{kope} proposed that the SMP in Ba(Fe$_{0.925}$Co$_{0.075}$)$_2$As$_2$ is associated with a vortex structural phase transition from rhombic to square lattice occuring at field and temperature corresponding to the minimum point of the magnetic-relaxation rate. In LiFeAs too, the SMP has been attributed to a vortex structural phase transition \cite{pramanik}. In BaFe$_{1.82}$Ni$_{0.18}$As$_2$, though a fishtail effect appears in the $M(H)$ curve, it is not associated with the crossover in a vortex pinning regime within the collective pinning scenario \cite{salem1}. On the other hand, in Ca(Fe$_{1-x}$Co$_{x}$)$_2$As$_2$ the SMP is absent and the analysis of temperature- and field-dependent magnetic relaxation data suggests that the vortex dynamics is consistent with the plastic creeping model \cite{pramanik1}. In contrast to the above, there are only few studies on vortex dynamics in the 1111 family of $R$FeAs(O,F) ($R$ = La, Ce, Pr, Nd, Sm etc.) compounds. Investigation of vortex dynamics by magnetization-relaxation measurements in SmFeAsO$_{0.9}$F$_{0.1}$ \cite{yang} and NdFeAsO$_{0.9}$F$_{0.1}$ \cite{proz1,moore} suggests that the vortex behavior is consistent with weak collective pinning and creep. Local measurements by magneto-optical imaging and microscopic Hall sensors in underdoped PrFeAsO$_{0.9}$ and NdFeAsO$_{0.9}$F$_{0.1}$ indicate that the pinning is due to the collective creep and the origin of the SMP is due to a order-disorder transition of the vortex lattice \cite{beek}.\\

As the vortex dynamics of the pnictide superconductors is sensitive to doping, anisotropy, etc. one has to study each system separately as a function of these parameters in order to understand the origin of the SMP. In this report, we have investigated the vortex dynamics and origin of SMP in the optimally-doped PrFeAsO$_{0.60}$F$_{0.12}$ sample by dc magnetization measurements, and temperature and field dependence of dynamic magnetization-relaxation rate $Q$. Data have been analyzed by the vortex collective pinning model and the method of generalized inversion scheme (GIS) \cite{schnack,wen}. Analysis of the temperature and field dependence of magnetic relaxation suggests that the vortex dynamics can be described by the collective pinning and creep, and a crossover from the elastic to plastic creep regime with increasing temperature and magnetic field. \\

\section{Experimental technique}
The polycrystalline sample used in this study was prepared by solid-state reaction method. The details of the sample preparation, powder x-ray diffraction analysis and magnetic characterization have been discussed in our earlier reports \cite{bhoi1,bhoi}. The phase purity of the sample was examined from x-ray and energy dispersive x-ray analysis \cite{bhoi1,bhoi}. Both the zero field cooled (ZFC) and field cooled susceptibilities start to deviate from the normal behavior below 48 K due to the appearance of a diamagnetic signal, which is close to the zero-resistance temperature \cite{bhoi1}. At 4 K, the shielding and Meissner fractions are calculated to be 88\% and 35\%, respectively, from the dc susceptibility data at $H$ = 10 Oe. The shielding fraction of the sample is $\sim$ 100\% at 4 K determined from the real part of the ac susceptibility curve at $H_{ac}$ = 3 Oe. The dynamic magnetization-relaxation measurements were carried out in a VSM-PPMS (Quantum Design) using the following protocol. The sample was cooled down to the desired temperature in the ZFC mode and then the magnetic field is swept and we measured the magnetic moment following the routes: 0 $\rightarrow H_{max} \rightarrow $ $-$1 T $\rightarrow$ 0, with different field sweeping rates $dH/dt$ = 40, 80 and 160 Oe/s. The magnetization-relaxation rate $Q$ is defined as
\begin{equation}
Q \equiv \frac{d \ln j_s}{d \ln(dH/dt)} \equiv \frac{d \ln(\Delta M)}{d \ln(dH/dt)},
\label{Q}
\end{equation}
where $j_s$ is the transient superconducting current density and $\Delta M=(M_+-M_-)$, $M_+(M_-)$ is the branch of the magnetization for $dH/dt$ $<$0 ($dH/dt$ $>$ 0).
\begin{figure}[!htbp]
\includegraphics[width=0.9\textwidth]{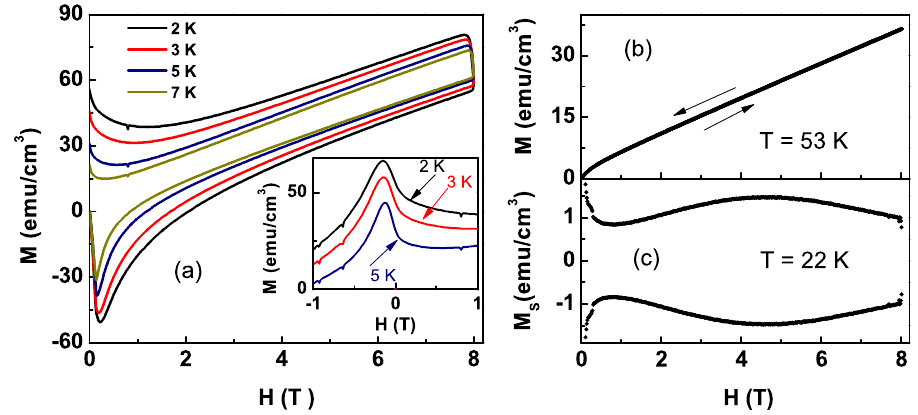}
\caption{(color online) (a) Magnetization hysteresis $M(H)$ loop of the PrFeAsO$_{0.60}$F$_{0.12}$ sample at different temperatures. Inset: the field-decreasing branch of hysteresis loops at 2, 3 and 5 K showing the effect of flux jump in the form of small kinks due to the thermomagnetic instability. (b) $M(H)$ loop at 53 K, just above the superconducting transition temperature. Arrows indicate the direction of increasing and decreasing magnetic field. (c) The superconducting irreversible signal $M_s(H)$ at 22 K exhibiting the second magnetization peak. $M_s(H)$ curve is obtained by subtracting the paramagnetic component $M_p = (M_{+} + M_{-})/2$, where $M_+(M_-)$ is the branch of the magnetization for $dH/dt<0(dH/dt>0)$, from the measured $M(H)$ loop.}\label{M(H)}
\end{figure}
\section{Results and discussion}
\subsection{Magnetization-hysteresis loop and dynamic magnetic relaxation}
\begin{figure}[!htbp]
\includegraphics[width=0.8\textwidth]{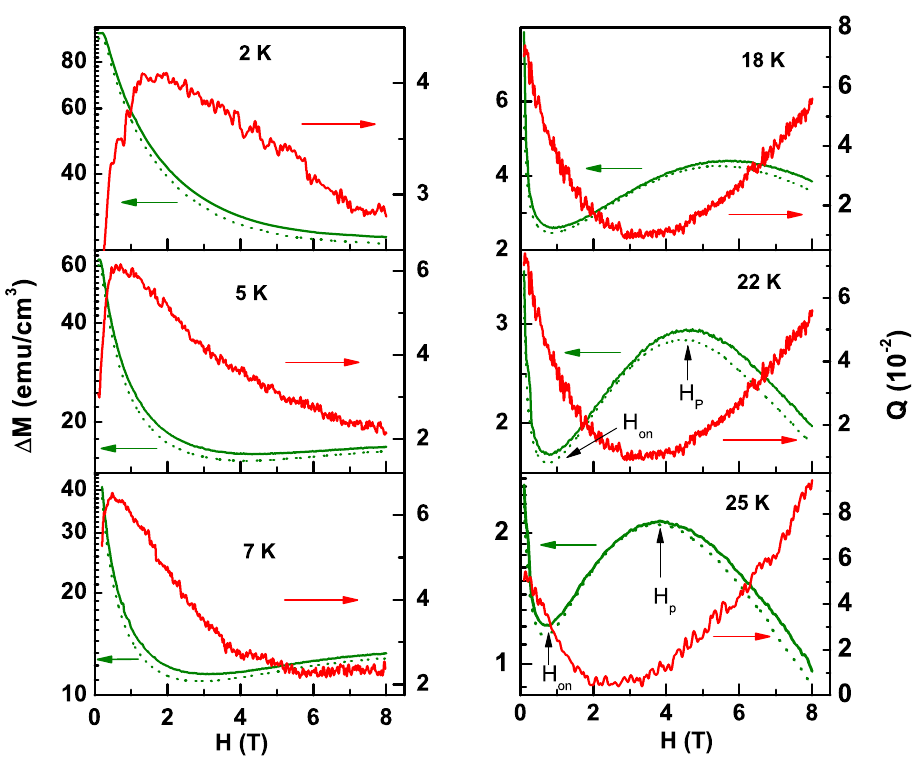}
\caption{(color online) Field dependence of $\Delta M$ at different temperatures with field sweeping rate 160 Oe/s (solid line) and 40 Oe/s (dotted line) and the dynamic magnetization-relaxation rate $Q$ (red line). Left axis shows the magnetization and right axis is for the dynamic magnetization-relaxation rate. Arrows indicate the onset field, $H_{on}$, and the peak field, $H_p$, of the second magnetization peak.}
\label{Q(H,T)}
\end{figure}
Figure \ref{M(H)} shows the magnetization hysteresis loops $M(H)$ of the PrFeAsO$_{0.60}$F$_{0.12}$ sample in the superconducting and normal state (just above $T_c$). One can clearly see that the superconducting hysteresis loop arises from the flux gradient produced by the pinning of flux lines. A small section of the field-decreasing branch of hysteresis loops at 2, 3 and 5 K is shown in the inset of Fig. \ref{M(H)}(a), wherein the signature of flux jumps is visible in the form of small kinks. Usually, flux jumps in superconductors occur in the low-field and low-temperature regime due to the effect of thermomagnetic instability (TMI) \cite{johnse}. Due to the thicker sample used in our study, the effect of flux jump is not strong enough though it is detectable as shown in the inset of Fig. \ref{M(H)}(a). The observed $M(H)$ curve in the superconducting state is the sum of a superconducting irreversible signal ($M_s$) and a paramagnetic component ($M_p$). Analysis of dc susceptibility, $\chi(T)$, in the normal state shows that the paramagnetic component of magnetization comes from the Pr$^{3+}$ magnetic moments \cite{bhoi}. As the sample has a large magnetic background in the normal state, we have measured magnetic hysteresis loop at 53 K, just above $T_c$ as shown in Fig. \ref{M(H)}(b). Figure \ref{M(H)}(b) shows no observable hysteresis, thereby excluding the presence of irreversible magnetization in the sample due to the free iron nanoparticle or magnetite traces. Figure \ref{M(H)}(c) shows that the superconducting irreversible signal $M_s(H)$ at 22 K exhibits a second magnetization peak. $M_s(H)$ curve is obtained by subtracting the paramagnetic component $M_p = (M_{+} + M_{-})/2$ from the measured $M_+$ and $M_-$ branches of $M(H)$ loop. In a polycrystalline sample, the gap $\Delta M$ in the magnetization loop can be split into intergranular (global) and intragranular (local) parts \cite{clem}. In the low-magnetic field region, $\Delta M$ is predominantly caused by the intergranular current, but in the high-field region $\Delta M$ results largely due to the intragranular current. This has been confirmed from the magnetization loop measurements on the bulk and powder sample in cuprates and pnictides \cite{mull,chen}. From the study of remnant magnetization as a function of maximum applied field in Sm and Nd iron oxypnictides, Yamamoto \emph{et al.} \cite{yama} showed that the intergrain current persists only up to a few hundred Oe. Since our study goes beyond this low-field region, we may assume that $\Delta M$ originates mainly from the intragrain current density.\\

In order to investigate the mechanism of vortex pinning and origin of the SMP in PrFeAsO$_{0.6}$F$_{0.12}$ sample, we have followed the dynamic magnetic-relaxation technique. Figure \ref{Q(H,T)} shows the field dependence of $\Delta M$ measured with the field sweeping rates of 160 and 40 Oe/s, and the value of $Q$ at different temperatures was calculated using equation (\ref{Q}). The arrows in the figure point to the characteristic fields: the onset field, $H_{on}$, and the peak field, $H_p$, of SMP. The difference between $\Delta M$ measured at 160 and 40 Oe/s can easily be distinguished, which indicates a giant vortex creep as observed in the cuprate superconductors. It is also clear from the figure that both $H_{on}$ and $H_p$ are dependent on the magnetic field sweeping rate and their values decrease with the decrease in field sweeping rate. Similar vortex creep has also been observed in single crystals of Ba(Fe$_{1-x}$Co$_{x}$)$_2$As$_2$ \cite{shen,proz} and Ba$_{1-x}$K$_{x}$Fe$_2$As$_2$ \cite{salem}, and polycrystalline SmFeAsO$_{0.9}$F$_{0.1}$ sample \cite{yang}. The value of $Q$ at 1 T and 7 K is as large as 6\% which is comparable to that observed in cuprate superconductors (e.g., 4\% in YBa$_2$Cu$_3$O$_7$ \cite{yesh}) but one order of magnitude larger than that for MgB$_2$ \cite{wen2}. There are several important features that can be identified from the field dependence of $Q$: (1) $H_{on}$ is weakly temperature dependent in the high-$T$ region but increases rapidly with decreasing temperature exceeding the limit of our measuring range (8 T) below 5 K. (2) In the low-temperature region (left panel of Fig. \ref{Q(H,T)}), a peak in $Q(H)$ is observed where the field dependence of $\Delta M$ changes its slope rapidly and the peak position shifts to lower field with increasing temperature. The low-field peak in the $Q(H)$ curve has also been observed in Ba(Fe$_{1-x}$Co$_{x}$)$_2$As$_2$ single crystal and has been explained as the signature of a crossover between two different regimes of vortex dynamics \cite{shen}. (3) In the high-temperature region (right panel of Fig. \ref{Q(H,T)}), there is a minimum in the $Q(H)$ curve which falls in between the fields $H_{on}$ and $H_p$. However, in case of Ba(Fe$_{1-x}$Co$_{x}$)$_2$As$_2$ single crystal this minimum roughly matches with the second peak position, $H_p$, in the $J(H)$ curve \cite{shen}.\\

\begin{figure}[!htbp]
\includegraphics[width=0.5\textwidth]{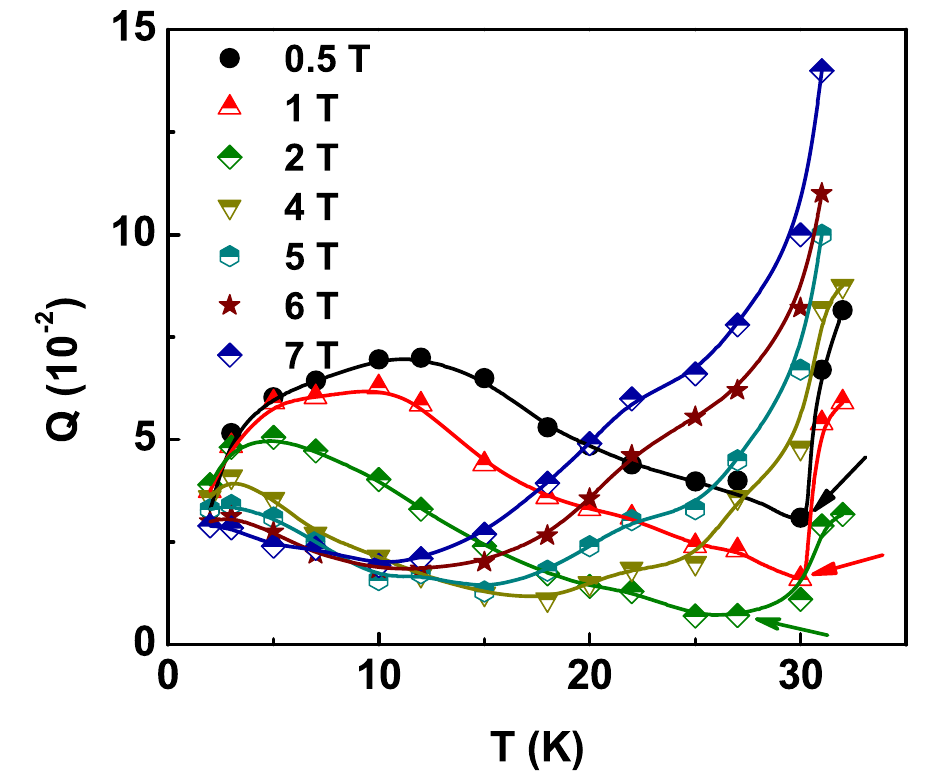}
\caption{(color online) Temperature dependence of the dynamic magnetization-relaxation rate $Q$ in different magnetic fields. Arrows indicate the temperature $T_m$ at which $Q(T)$ shows a minimum.}\label{Q(T)}
\end{figure}

Figure \ref{Q(T)} shows the temperature dependence of $Q$ for different magnetic fields. $Q$ shows a bell-like or bump-like shape in the intermediate-temperature region for low field. With increasing magnetic field, the bump-like feature shifts toward lower temperature and disappears above 4 T. For all magnetic fields, $Q(T)$ passes through a minimum at a temperature $T_m$ as indicated by arrows in Fig. \ref{Q(T)}. At 0.5 T, $T_m\approx$ 30 K ($\simeq$ 0.63 $T_c$) and with increase of magnetic field $T_m$ shifts toward lower temperature. Similar bump-like shape in low field has also been observed in the $Q(T)$ data of Ba(Fe$_{0.92}$Co$_{0.08}$)$_2$As$_2$ single crystal \cite{shen} and in the remanent magnetization relaxation data of Ba(Fe$_{0.93}$Co$_{0.07}$)$_2$As$_2$ \cite{proz} and Ca$_{0.25}$Na$_{0.75}$Fe$_2$As$_2$ \cite{habe} single crystals. The bump-like shape disappears and a plateau appears in $Q(T)$ or $S(T)$ for $H$ above 1 T and 3 T for Ca$_{0.25}$Na$_{0.75}$Fe$_2$As$_2$ \cite{habe} and Ba(Fe$_{0.92}$Co$_{0.08}$)$_2$As$_2$ \cite{shen}, respectively. In all these cases, this behavior has been attributed to the effect of collective pinning of vortices.
\subsection{Analysis based on the vortex collective pinning model}
\begin{figure}[!htbp]
\includegraphics[width=0.9\textwidth]{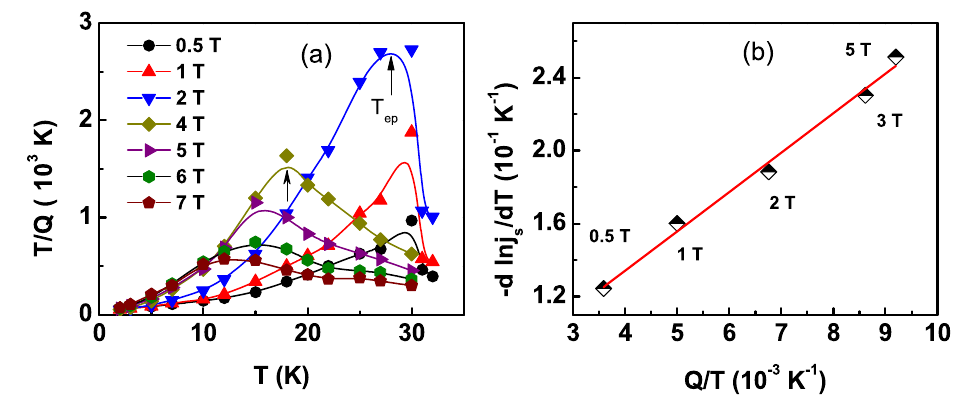}
\caption{(color online) (a) Temperature dependence of the $T/Q$ in different magnetic fields. The elastic to plastic crossover temperature, $T_{ep}$, was determined from the peak position of the $T/Q$ vs $T$ curve for different magnetic fields. Arrows indicate $T_{ep}$ for 2 and 4 T. (b) -$d$ln$j_s$/$d$T vs $Q/T$ at 7 K in different magnetic fields. The best fit line (red) has slope $C$ = 21.62$\pm$1.2.}\label{Q/T}
\end{figure}
In a type-II superconductor, the vortices normally move over the effective pinning barrier $U(j_s,T,H_e)$ by thermal activation with an average velocity $\bar{v} = v_{0}\exp\left(-\frac{U(j_s,T,H_e)}{k_BT}\right)$, where $v_0$ is the attempt hopping velocity and $H_e$ is the actual local magnetic induction. The electric field, $E$, induced by this vortex motion is given by \cite{ander},
\begin{equation}
E = v_{0}H \exp \left(-\frac{U(j_s,T,H_e)}{k_BT}\right).
\label{tafm}
\end{equation}
In a system with randomly distributed weak pinning centers, the current dependent effective pinning barrier can be written in a general form \cite{male}
\begin{equation}
U(j_s,T,H_e) = \frac{U_c(T,H_e)}{\mu(T,H_e)}\left[\left(\frac{j_c(T,H_e)}{j_s(T,H_e)}\right)^{\mu(T,H_e)}-1\right],
\label{U(J)}
\end{equation}
where $\mu$, $U_c$, and $j_c$ are the glassy exponent, intrinsic characteristic pinning energy, and the unrelaxed critical current density, respectively. According to the collective pinning theory, for a three-dimensional lattice the values of exponent $\mu$ are 1/7, 3/2 and 7/9 for the single vortex, small bundles and large bundles of vortex motion, respectively \cite{vino}. For $\mu$ = $-$1, equation (\ref{U(J)}) describes the Kim-Anderson model \cite{ander} and in the limit $\mu\rightarrow$0, it reduces to the Zeldov logarithmic model \cite{zel}. Any value of $\mu$, positive, negative or zero is physically meaningful. A negative $\mu$ corresponds to a finite dissipation in the small current limit and plastic vortex motion, while a positive $\mu$ corresponds to a vanishing dissipation in the small current limit and elastic vortex motion \cite{wen1}. From the general equations (1)-(\ref{U(J)}), Wen \emph{et al} \cite{wen} derived the following equation:
\begin{equation}
\frac{T}{Q(T,H_e)} = \frac{U_c(T,H_e)}{k_B} + \mu(T,H_e)CT
\label{T/Q}
\end{equation}
where $C =\ln\left[2v_0H_e/l(dH_e/dt)\right]$ is a parameter that is weakly temperature dependent and $l$ is the lateral dimension of the sample. Figure \ref{Q/T}(a) shows the $T/Q$ vs $T$ curves at different magnetic fields. The $T/Q$ curves go through a peak, the position of which varies with the applied field. The left side of the peak denotes a positive slope of the curve which corresponds to a positive $\mu$ [equation \ref{T/Q}] $-$ provided $U_c(T)$ has no strong temperature dependence $-$ and thereby elastic vortex motion while the right side of the peak has a negative slope which suggests a plastic vortex motion. Thus, as the strength of the applied field is increased, the low-temperature region over which the elastic vortex motion prevails is shortened. By extrapolating the curve $T/Q$ down to zero temperature, we have deduced the value of $U_c(0)$ at 0.5 T from Fig. \ref{Q/T}(a). The obtained value of $U_c(0)/k_B$ is about 36 K, which is comparable to those for SmFeAsO$_{0.9}$F$_{0.1}$ ($U_c(0)\sim$ 40 K) \cite{yang} and Ba(Fe$_{1-x}$Co$_{x}$)$_2$As$_2$ ($U_c(0)\sim$ 98 K) \cite{shen}. However, these values are order of magnitude smaller compared to $U_c(0)\sim$ 300 K for YBCO thin films \cite{wen} and beyond 3000 K for MgB$_2$ \cite{jin}, implying a quite small characteristic pinning energy for the pnictide superconductors. Assuming $U_c(T)$ is not a strong temperature-dependent function, the value of $\mu C$ can be obtained from the slope of the $T/Q$ vs $T$ curve. At $H$ = 0.5 T, $\mu C$ is evaluated to be 11.57$\pm$0.64. To estimate $C$, one can make use of the equation \cite{wen}
\begin{equation}
-\frac{d \ln j_s}{d T} = - \frac{d \ln j_c}{d T} + C\frac{Q}{T}
\label{C}
\end{equation}
which is valid in the temperature region where $j_s$ is not affected by the quantum creep at low temperatures and ln$j_s$ follows an almost linear decrease with $T$. $-d$ln$j_s$/$dT$ vs. $Q(T)/T$ has been plotted at 7 K for five different magnetic fields as shown in Fig. \ref{Q/T}(b). The slope of the best-fit straight line provides the value of $C$ = 21.62 $\pm$ 1.2. The parameter $C$ can also be evaluated from the equation \cite{schnack}
\begin{equation}
C = \lim_{T \to 0} \frac{-1}{Q}\frac{d\ln j_s}{d\ln T}
\end{equation}
The extrapolation of $(-1/Q)(d\ln j_s/d\ln T)$ vs. $T$ curve to 0 K gives $C$ = 21.8. The values of $C$ obtained from these two procedures are close to each other and slightly smaller than that reported for Ba(Fe$_{1-x}$Co$_{x}$)$_2$As$_2$ single crystals ($C$= 28.56) \cite{shen} but close to that for YBa$_2$Cu$_4$O$_8$ ($C$= 19) \cite{wen}. From the values of $\mu C$ and $C$, we have estimated the value of $\mu$ to be 0.53. The positive value of $\mu$ indicates an elastic vortex motion. In the later section, we will show that this value of $\mu$ is consistent with that estimated from the generalized inversion scheme.\\

\subsection{Analysis based on generalized inversion scheme}
\begin{figure}[!htbp]
\includegraphics[width=0.9\textwidth]{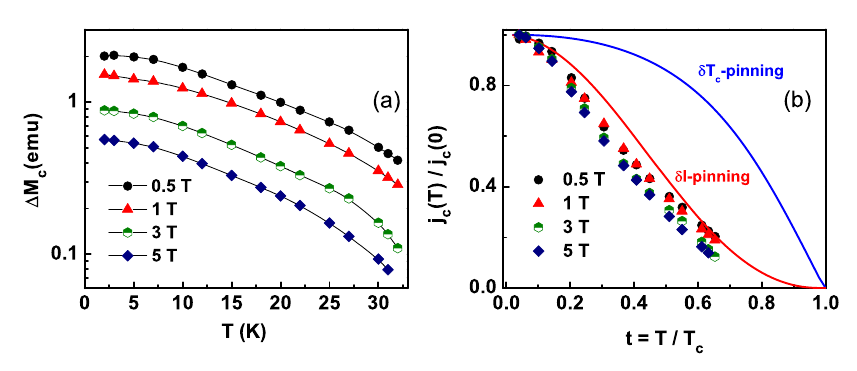}
\caption{(Color online)(a) Temperature and field dependence of the unrelaxed $\Delta M_c(T,H_e)$ evaluated by following GIS. (b) Temperature dependence of $j_c(T)/j_c(0)$ in different magnetic fields along with the theoretical prediction for $\delta l$ and $\delta T_c$ pinning.}\label{jcT}
\end{figure}
\begin{figure}
\includegraphics[width=0.5\textwidth]{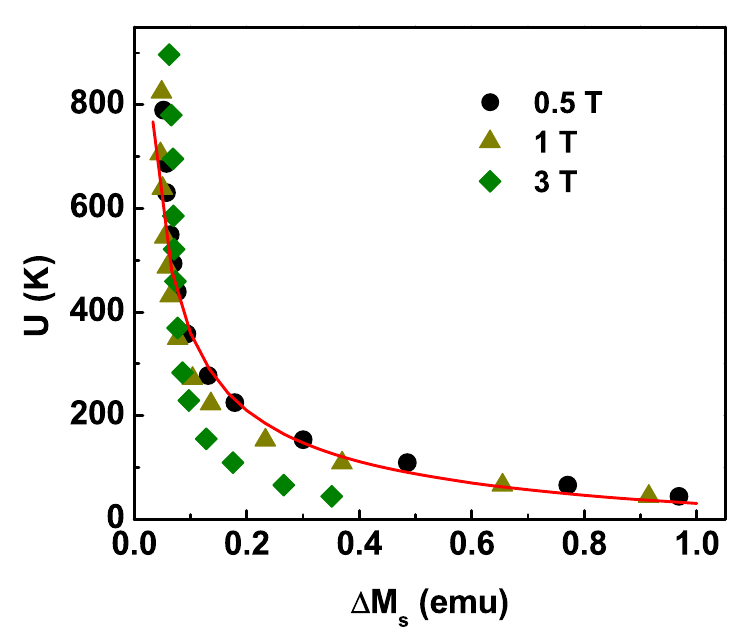}
\caption{(color online) Current dependence of the effective pinning energy $U(T,H_e)$ in different magnetic fields evaluated by following the method of GIS. The solid line is a theoretical fit of equation (\ref{U(J)}) to the $U(j_s,T)$ data for $H$ = 0.5 T in the temperature region 2-20 K. Using the obtained best fit parameters $U_c(0)$ = 48 K and $\mu$ = 0.59 this curve has been extended for $T >$ 20 K.}\label{UcT}
\end{figure}

In order to evaluate the temperature and field dependence of the unrelaxed critical current density $j_c(T,H_e)$ and the corresponding characteristic pinning energy $U_c(j_s,T,H_e)$ directly from relaxation data, we have followed the generalized inversion scheme proposed by Schnack \emph{et al} \cite{schnack} and Wen \emph{et al} \cite{wen}. Though the present sample is polycrystalline in nature, scanning electron microscope images reveal large, oriented, and well-connected plate-like crystallites. The average grain size in this sample is larger than those normally reported in oxypnictide polycrystalline samples and the plate-like grains have a tendency to orient their $ab$-plane along the broad face of the sample \cite{bhoi}. All these led us to venture for the GIS analysis on the relaxation data of our polycrystalline sample. The GIS scheme is more general as it does not require a priori assumption about the explicit temperature or field dependence of $U$. The basic postulates of GIS are: (1) thermally activated flux motion can be described by equation (\ref{tafm}); (2) the activation energy can be expressed as product of two general functions $g(t = T/T_c,H_e)$ and $f\left[j_s(T, H_e)/j_c(T, H_e),H_e\right]$, i.e., $U(j_s,T,H_e)$ = $U_c(0,H_e)\times g(t,H_e)\times f\left[j_s(T, H_e)/j_c(T, H_e), H_e\right]$ with $f(1, H_e)$ = 0, since $U(j_c,T,H_e)$ = 0 by definition of the critical current density $j_c$; and (3) $g(t,H_e)\propto [j_c(T, H_e)/j_c(0, H_e)]^pG(t)$, here $G(t)$ and $p$ depend on the specific pinning models \cite{wen}. According to GIS, the temperature dependence of the true critical current density $j_c$ at a particular field can be evaluated from the following integral \cite{schnack,wen}
\begin{equation}
j_c(T)=j_c(0)\times \exp\left[\int_0^T \frac{CQ(T')[1-\frac{d\ln G(T')}{d \ln T'}]+\frac{d\ln j_s(T')}{d\ln T'}}{1+pQ(T')C}\times\frac{dT'}{T'}\right],
\label{jT}
\end{equation}
where $j_c(0)$ is the true critical current density at 0 K. In the above equation, $Q(T)$ and $j_s(T)$ are the measured values.
In order to apply the above procedure, one must know the value of $p$ as well as the function $G(t)$. For a two-dimensional pancake system, $p$ = 1 and $G(t) = \sqrt{(1+t^2)/(1-t^2)}$ with $t = T/T_c$. Similarly, for a three-dimensional single vortex, $p$ = 0.5 and $G(t) = (1+t^2)^{5/4}/(1-t^2)^{1/4}$ (Ref.\cite{wen}). In the underdoped PrFeAsO$_{0.9}$, it has been shown that in the low-field region the vortex pinning is in the three-dimensional single-vortex limit \cite{beek}. Therefore, we have taken $p$ = 0.5 and $C \sim$ 22 to calculate the values of $j_c(T)$ following the method of GIS. As the supercurrent density $j_s \propto \Delta M$, therefore, we have plotted $\Delta M_c(T)$ in Fig. \ref{jcT}(a) for different magnetic fields. In type II superconductors, the pinning originates from two basic mechanisms: due to the spatial fluctuation of superconducting transition temperature $T_c$ ($\delta T_c$-pinning) and due to the spatial variation in the mean free path $l$ near a lattice defect ($\delta l$-pinning). In the single vortex regime, the theoretical curve for $\delta l$-pinning is given by $j_c(T)/j_c(0)$=$(1+t^2)^{-1/2}(1-t^2)^{5/2}$ and that for the $\delta T_c$-pinning is given by $j_c(T)/j_c(0)$=$(1-t^2)^{7/6}(1+t^2)^{5/6}$ (Ref. \cite{blatt,gries}). In Fig. \ref{jcT}(b), we have shown the temperature dependence of experimentally derived $j_c(T)/j_c(0)$ values together with the theoretical predictions for two basic pinning mechanisms. It is clear from the figure that the experimentally determined values are close to the $\delta l$-pinning mechanism. This is also consistent with the prediction that the flux pinning in the charge-doped pnictides can be described by the mean-free path fluctuations introduced by the dopant atoms \cite{beek1}. Figure \ref{UcT} shows $U(j_s,T,H_e)$ obtained by GIS at three different fields for the PrFeAsO$_{0.60}$F$_{0.12}$ sample. From the fitting of equation (\ref{U(J)}) to the data for 0.5 T, the parameter $U_c(0)$ is evaluated to be 48 K which is close to the value ($\sim$36 K) determined from the analysis using collective pinning model. From the fit, we also get $\mu$ = 0.59, which is close to that obtained ($\mu$ = 0.53) from the $T/Q$ vs. $T$ curve. It should be noted that the values of $\mu$ determined here reflect just an averaged one, which, in principle, is also current dependent.
\section{Phase Diagram and second magnetization peak}
\begin{figure}[h]
\includegraphics[width=0.5\textwidth]{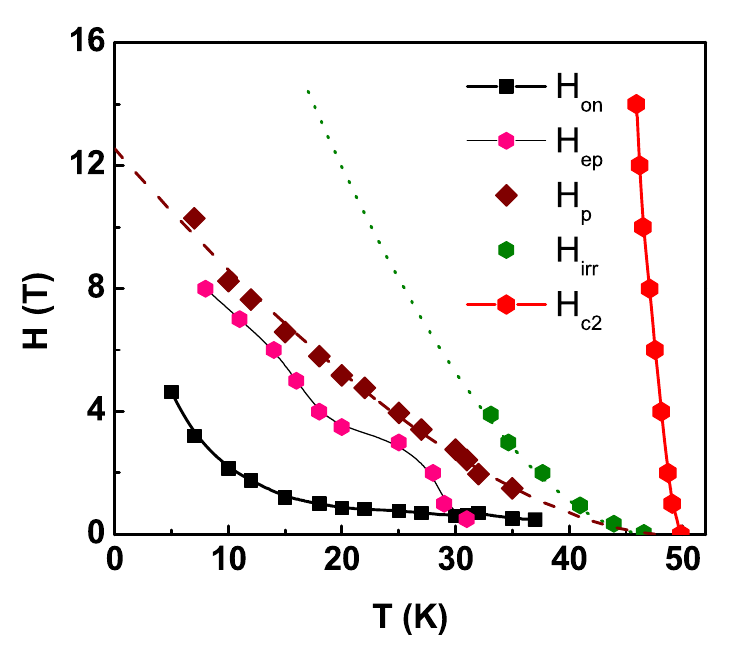}
\caption{(color online) Vortex phase diagram of PrFeAsO$_{0.60}$F$_{0.12}$ sample describing the temperature dependence of the onset field, $H_{on}$, the elastic to plastic creep crossover field, $H_{ep}$, the peak position $H_p$, the irreversibility field $H_{irr}$ and the upper critical field $H_{c2}$. Data for $H_{irr}$ and $H_{c2}$ are taken from Ref. \cite{bhoi} and \cite{bhoi1}, respectively. The dashed line and dotted line describe the power law fit $(1-\frac{T}{T_c})^n$ to $H_p$ and $H_{irr}$ data with $n$ = 1.6 for $H_p$  and 1.7 for $H_{irr}$.}\label{HT}
\end{figure}

In Fig. \ref{HT}, we have plotted the temperature dependence of the characteristic fields, viz., $H_{on}$, $H_p$, $H_{ep}$ (the elastic to plastic crossover field), $H_{irr}$ and $H_{c2}$. The $H_{ep}(T)$ values are obtained from the $T/Q$ vs. $T$ curves for different $H$ (Fig.\ref{Q/T}(a)). The $H_{ep}(T)$ curve is located in between the onset, $H_{on}$ and the peak, $H_p$ of the SMP. The characteristic fields $H_{on}$ and $H_p$ follow a concave-shaped decrease with the increase in temperature. The SMP has been observed in the conventional as well as high-temperature superconductors and various mechanisms have been proposed to explain the origin of the same. Foremost among them are (i) a change in the dynamics of the vortex lattice $-$ a crossover from the single-vortex pinning regime to the bundle pinning regime \cite{kel}, (ii) a change in the vortex creep mechanism $-$ a crossover from the elastic to  plastic creep \cite{blat}, (iii) a first-order phase transition from an ordered ``elastically pinned" low-field vortex phase, the so-called Bragg-glass \cite{gia} to a high-field disordered phase characterized by the presence of topological defects \cite{kir,mik} and (iv) a structural vortex lattice transition (from rhombohedral to square \cite{rose} or from triangular to square lattice \cite{gilard}) well within the Bragg glass regime.\\
For the present case, the absence of a mirror image correlation between $\Delta M$ and $Q$ rules out the possibility of a crossover from the single-vortex pinning regime to the bundle pinning regime as the origin of SMP \cite{kel}. The $T$ dependence of $T/Q$ curve for different $H$ (Fig. \ref{Q/T}) has unambiguously established the occurrence of a crossover in the flux dynamics, from elastic to plastic creep with increasing temperature. However, this crossover line, $H_{ep}$, does not coincide with any of the characteristic fields, viz.,$H_{on}$ or $H_p$ of the SMP; rather, it lies in between $H_{on}$ and $H_p$. Also, there is no anomalous feature in the $M(H)$ curve at $H_{ep}$. It may be mentioned that in YaBaCuO$_{7-\delta}$, $H_{ep}$ coincides with $H_p$ \cite{gil1,abu}, whereas in Nd$_{1.85}$Ce$_{0.15}$CuO$_{4-\delta}$ the creep crossover occurs well above $H_p$ \cite{gil}. In case of the crossover in the flux creep behavior occurring below $H_{p}(T)$, both $H_{p}(T)$ and $H_{irr}(T)$ are controlled by plastic pinning and their temperature dependence are expected to be of the form $(1-\frac{T}{T_c})^n$ \cite{shen,proz}. For Ba(Fe$_{1-x}$Co$_x$)$_2$As$_2$ ($x$ = 0.08) single crystal, $H_{p}(T)$ and $H_{irr}(T)$ lines follow such kind of temperature dependence with $n$ = 1.7 and 1.52, respectively \cite{shen}; for $x$ = 0.07, $H_{p}(T)$ goes with $n$ = 1.5 \cite{proz}. In the present case, we observed $n$ = 1.6 and 1.7 for the $H_{p}(T)$ and $H_{irr}(T)$ lines respectively, in agreement with the data for Ba-122 crystal.\\

It is well accepted that a crossover from elastic to plastic vortex creep may accompany a quasi-order-disorder (OD) transition of the vortex lattice. The vortex phase of the superconductors is determined by the competition between the energy of thermal fluctuations, the pinning energy generated by the quenched disorder, $E_p$, and the elastic energy of the vortex system, $E_{el}$ \cite{gil,gia,vino1,nishi}. $E_p$ and $E_{el}$ are directly related to the superconducting parameter such as penetration depth $\lambda$, coherence length $\xi$, pinning parameter and the anisotropy factor. At low $T$, where thermal energy is small compared to $E_p$ and $E_{el}$, the OD transition is roughly described by the equality \cite{gil,gil1}
\begin{equation}
E_p(T,H) = E_{el}(T,H).
\end{equation}
For static condition where no current flows in the sample, OD transition field is practically independent of $T$ in the low-$T$ region, far below $T_c$ where the superconducting parameters vary weakly with temperature \cite{gil,gil1,deli,khay}. In global magnetic measurements, the transition is relatively wide, and the transition field lies somewhere in between $H_{on}$ and $H_p$, though $H_{on}$ is usually considered as the transition field. In cuprates, the occurrence of SMP is commonly associated with this OD transition \cite{gil,gil1,deli,khay}. Unlike cuprates, in the present case $H_{on}$ exhibits a concave-shaped increase with the decrease of $T$, as observed for other characteristic fields such as, $H_{ep}$ and $H_p$. La$_{2-x}$Sr$_x$CuO$_4$, with $x$ = 0.126 and similar doping, exhibits a broad SMP with characteristics that are strongly temperature dependent down to low temperature \cite{rose,razd}. This behavior was explained by considering the softening of the vortex lattice associated with the square to rhombic vortex lattice transition as the source for the SMP \cite{kope,rose}.
Alternatively, the upward curvature in the $T$ dependence of $H_{on}$ and $H_p$ in the low-$T$ region may be explained as a dynamic effect caused by the finite current $j_s$ induced in the specimen during standard dc magnetization experiments, which reduces the effective pinning energy $U(j_s,T,H)$ \cite{miu}. In a crude approximation, in the dynamic condition, the appropriate energy balance relation should be
\begin{equation}
 U(j_s, T, H)\propto E_{el}(T,H).
 \label{uel}							
\end{equation}
With $U(j_s,T,H) \propto T$ and $E_{el}(T,H)\propto\lambda^{-2}H^{-1/2}$ (independent of $j_s$) one obtains a $T^{-2}$ dependence for $H_{on}$,
\begin{equation}
H_{on}(T)\propto\lambda^{-4}T^{-2}
\label{honT}
\end{equation}
at low temperature.
\begin{figure}
\includegraphics[width=0.9\textwidth]{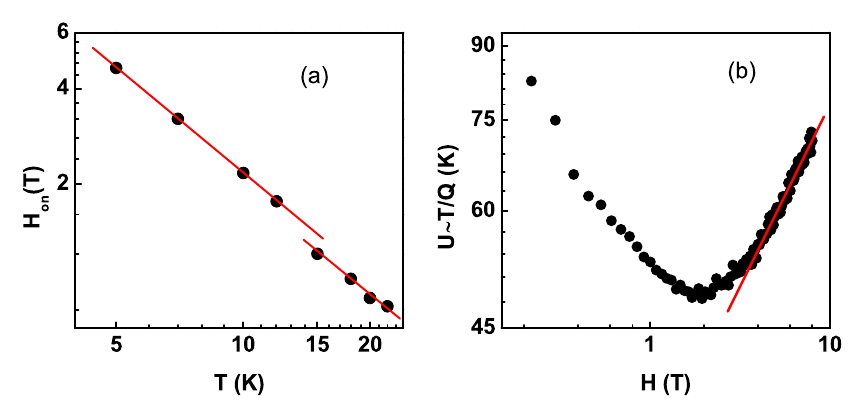}
\caption{(color online) (a) The low temperature variation of $H_{on}(T)$ for $T/T_c\leq$ 0.5 in double logarithmic scale. In the low-$T$ range, $H_{on}(T)\propto 1/T^{1.1}$. The solid lines represent a linear fit in two temperature intervals to emphasize the presence of a inflection-like behavior located between 12 and 15 K. (b) The variation of $U(\sim T/Q)$ with $H$ at 2 K on a log-log scale. Solid line describes the power law fit $U(H)\propto H^{0.4}$ for the $H$ domain corresponding to $H_{on}$ at $T/T_c \leq$ 0.5.}\label{hon}
\end{figure}
In the high temperature range ($0.5 < T/T_c < 1$), the dependence of superconducting parameters on $T$ is important to determine the origin of pinning.
In this region, for $\delta l$ pinning, $H_{on}(T)$ is expected to show a peak and follow the expression $H_{on}(0)\left[1-(T/T_c)^4\right]^{-1/2}$ below the peak \cite{gil1}. Indeed, we have observed a weak peak at $T \sim$ 0.67$T_c$ which may be related to the $\delta l$-pinning of the vortices. However, due to limited data points in a narrow temperature range, fitting could not be done unambiguously.\\

For the low-$T$ range ($T/T_c <$ 0.5), $H_{on}(T)$ plotted in a double logarithmic scale is shown in Fig. \ref{hon}(a). In this region, $H_{on}$ varies as $T^{-1.11}$ which is smaller than the exponent predicted in equation (\ref{honT}). In FeSe$_{1-x}$Te$_x$, such a low exponent was attributed to the increase of $U$ with increasing $H$, specific for the collective creep regime \cite{miu2}. From a plot of $U(\sim T/Q)$ vs $H$ at 2 K (Fig. \ref{hon}(b)) we observed a $H^{0.4}$ dependence of $U(H)$ in the $H$ domain corresponding to $H_{on}$ and at $T/T_c \leq$ 0.5, indicating $U(T,H)$ varies as $TH^{0.4}$. Combining this with equation (\ref{uel}), and neglecting the $T$ dependence of $\lambda$ in the low-$T$ limit, one obtains $H_{on}(T)\propto T^{-1.11}$, as observed in Fig. \ref{hon}(a). The observed upturn in $U(H)$ below 2 T (Fig. \ref{hon}(b)) may be attributed to the the effect of TMI appearing in the low-$T$ domain \cite{miu2}. There is another striking similarity of our result with those for FeSe$_{1-x}$Te$_x$ \cite{miu2} and La$_{2-x}$Sr$_x$CuO$_4$ crystals \cite{miu}; we find an inflection-like point in the $H_{on}$ vs $T$ behavior between 12 and 15 K (Fig. \ref{hon}(a)). The location of this inflection point is found to be in good agreement with the results obtained for $\lambda(T)$ in underdoped PrFeAsO$_{1-y}$ single crystal \cite{hashi} and following Miu \emph{et al.} \cite{miu,miu2} we associate this feature with the $T$ dependence of the superfluid density ($n_s\propto \lambda^{-2}$) in the case of two-band superconductivity.
\section{Conclusion}
In summary, we have investigated the vortex dynamics of the optimally-doped PrFeAsO$_{0.60}$F$_{0.12}$ sample by dc magnetization and dynamic magnetization-relaxation measurements. The field dependence of the superconducting irreversible magnetization reveals a second magnetization peak. In low-field and in the intermediate-temperature region, $Q(T)$ exhibits a bell-like or bump-like shape. In this region, data analysis based on the vortex collective pinning model and GIS suggests that the vortex dynamics can be better described by elastic vortex motion. Analysis of the temperature and field dependence of $Q$ suggests a crossover of the vortex dynamics from the elastic to plastic creep regime with increasing temperature and magnetic field. The temperature dependence of the critical current density is consistent with the pinning due to the spatial variation in the mean free path near a lattice defect ($\delta l$-pinning). Analysis of present data persuades for an order-disorder like transition of the vortex phase in our sample followed by a crossover from elastic creep to plastic creep in the vortex motion.

\section{Acknowledgement}
The authors would like to thank Prof. S. Das, N. Khan, A. Midya and A. Paul for technical help during measurements.

\end{document}